\documentclass[twocolumn, prl, amssymb,  aps, showpacs,preprintnumbers,
amsmath,showkeys,floatfix]{revtex4}

\usepackage{epstopdf}
\usepackage{graphics}
\usepackage{graphicx}
\usepackage{dcolumn}
\usepackage{bm}
\usepackage{longtable}
\usepackage{epsfig}
\usepackage{times}
\usepackage{url}
\usepackage{color}

\begin{document}
\title{All-Electromagnetic Control of Broadband Quantum Excitations Using Gradient Photon Echoes}

\author{Wen-Te \surname{Liao}}
\email{Liao@mpi-hd.mpg.de}

\author{Christoph H. \surname{Keitel}}

\author{Adriana \surname{P\'alffy}}
\email{Palffy@mpi-hd.mpg.de}

\affiliation{Max-Planck-Institut f\"ur Kernphysik, Saupfercheckweg 1, D-69117 Heidelberg, Germany}
\date{\today}
\begin{abstract}

A broadband photon echo effect in a three level $\varLambda$-type system interacting with two laser fields is investigated theoretically. Inspired by the emerging field of nuclear quantum optics which typically deals with very narrow resonances, 
we consider broadband  probe pulses that couple to the system in the presence of an  inhomogeneous control field. We show that such a setup provides an all-electromagnetic-field solution to implement high bandwidth photon echoes,  which are easy to control, store and shape on a short time scale and therefore may speed up future photonic information processing. The time compression of the echo signal and possible 
 applications for quantum memories  are discussed.

\end{abstract}

\maketitle

Manipulation and control of broadband quantum excitations on different time scales are an important task for fundamental and applied physics and  information technology. High bandwidths would allow controlling temporally short light pulses and therefore fast photonic information processing \cite{Reim2010}. However, a tradeoff between the allowed operation time and the system bandwidth plagues quantum transitions in atoms:
limited linewidths lead to information loss, broad ones on the other hand to very short coherence times. A radically different approach emerges when considering atomic nuclei which have very long coherence times and tiny bandwidths, automatically transforming any incoming pulse into a broadband excitation.
Inspired by the emerging field of nuclear quantum optics with broadband x-ray pulses \cite{Shvydko1996,Palffy2009,Liao2012a,Liao2012b,Heeg2013,Liao2014,Liao2014b,Vagizov2014}, we put forward 
how to  ease the dilemma above with an all-electromagnetic gradient photon echo setup
that can store broadband light pulses  in a novel and temporally scalable setup. 

\begin{figure}[b]
\vspace{-0.4cm}
  \includegraphics[width=0.48\textwidth]{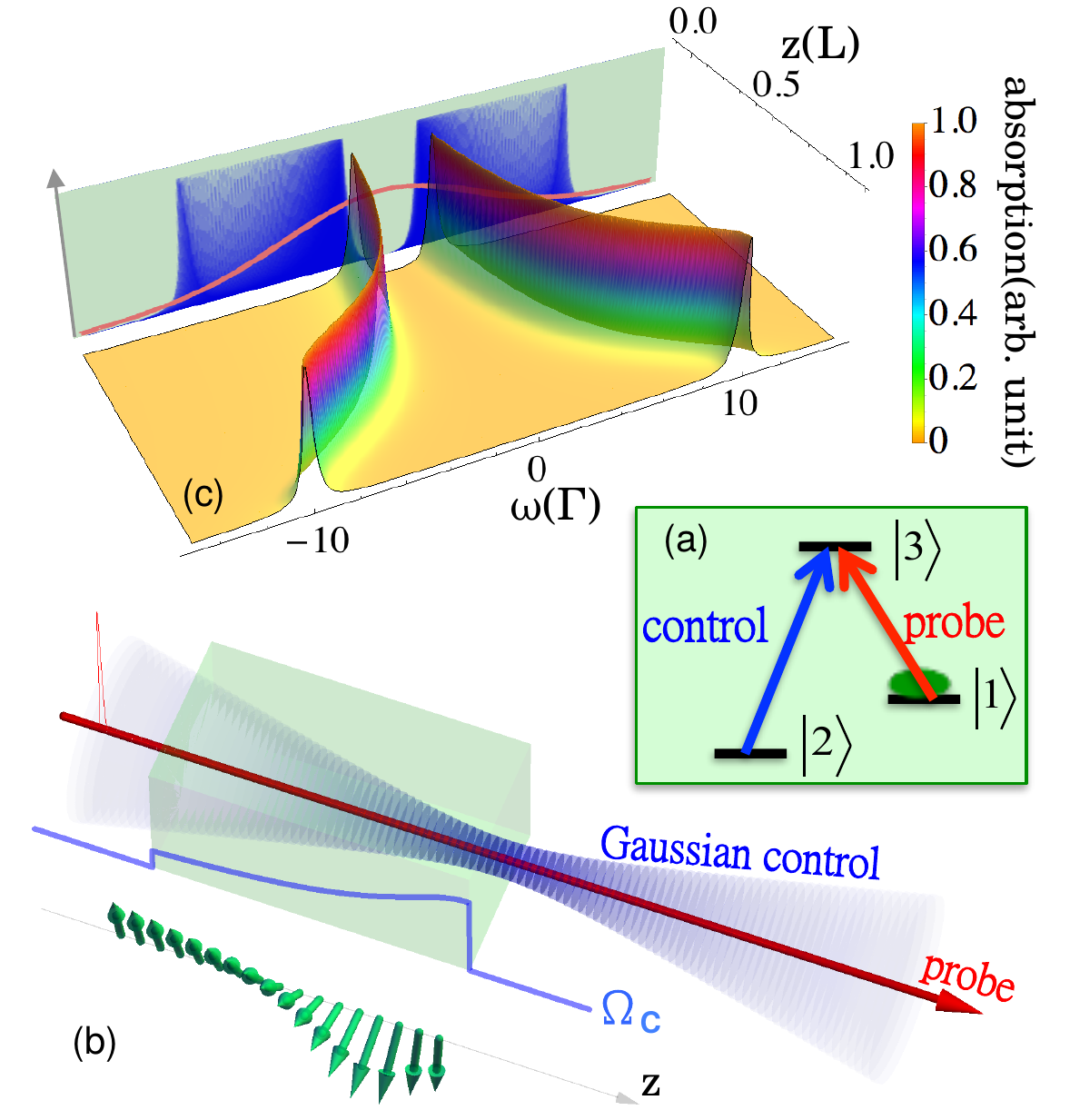}
  \caption{\label{fig1}
(Color online). (a) $\varLambda$-type three level  system. The blue (red) arrow depicts the control (probe) laser field which drives the transition $|2\rangle \rightarrow |3\rangle$ ($|1\rangle \rightarrow |3\rangle$). (b) All-electromagnetic photon echo setup. Blue translucent Gaussian beam depicts the spatial profile of the control field. The control Rabi frequency $\Omega_c$ is displayed by the blue raising curve on the side of the green cube. The array of short green arrows shows the position-dependent evolution of $\rho_{21}$ along the $z$-axis. (c) Position-dependent ATS. The energy splitting becomes larger due to the focusing of the control field, as shown by the projection of the ATS at diferent positions $z$ on the vertical plane.  The red Gaussian curve depicts the Fourier transform of a broadband incident probe pulse.
  }
\end{figure}

Gradient photon echoes are typically obtained in two- or three-level systems by introducing a frequency shift gradient in the sample which is reversed at some instant in time. The rephasing of the dipoles in the sample will lead to the emission of a photon echo \cite{Hetet2008}. The generic system considered here is a three-level $\varLambda$-type configuration as depicted in Fig.~\ref{fig1}(a).
Two fields, denoted  by ``probe'' and ``control'' couple the three levels in a setup reminding of electromagnetically induced transparency (EIT) \cite{Hau1999,Kocharovskaya2001,Liu2001,Phillips2001,Longdell2005,Gorshkov2007b,Gorshkov2007prl,Lin2009,Heinze2013,Chen2013}. The gradient, usually introduced by an additional magnetic or electric field, arises in the all-electromagnetic  scheme  from a spatial intensity profile of the continuous wave control field as discussed in Ref. \cite{Sparkes2010} and illustrated here in Fig.~\ref{fig1}(b). 
Two novel features are introduced in our setup: (i) the echo is produced by a phase flip of the control field,  (ii) the probe field bandwidth is larger than the  transparency window induced by the control field, as illustrated in Fig.~\ref{fig1}(c). The Rabi frequency of the latter is also larger than the spontaneous decay rate. In this region the so-called Autler-Townes splitting (ATS) comes into play \cite{Autler-Townes,Anisimov2011,Liao2012b}. This is the opposite case to the typical EIT scenario and was so far never investigated. 
We show that our setup offers new possibilities to store and shape ultra-short pulses by multi-mode interference. 
Compared to other mechanisms such as the far-off-resonance Raman transitions \cite{Reim2010,Reim2011}, our scheme may ease the need of a broadband read-write field for light storage at large bandwidths \cite{deRiedmatten2010}.
As another new feature, we find that the time scaling symmetry of this setup provides means to control the echo pulse duration on different time scales. This is of major 
relevance for both the production and manipulation of ultra-short pulses \cite{Radeonychev2010,Antonov2013,Antonov2013b} and for gradient echo memory systems \cite{Hetet2008b,Hosseini2009,Sparkes2010,Sparkes2012,Sparkes2013}.

We start with a simplified case involving a uniform control field.  Initially, the entire population is  situated in the ground state $|1\rangle$.
The transition $|1\rangle \rightarrow |3\rangle$ ($|2\rangle \rightarrow |3\rangle$) is driven by a weak ultra-short probe laser pulse (strong continuous-wave control field).  
We consider here a generic three-level system with an equal branching ratio characterized by the lifetime $\tau$ of the upper level $|3\rangle$, which is used as time unit for the presented  analysis. Motivated by fast quantum-excitation control \cite{Reim2010}, the time scale of interest here is much shorter than $\tau$, preventing spontaneous decay noise.

To describe the dynamics of the system, we use the Maxwell-Bloch equations \cite{Scully2006} in the region of $\vert\Omega_{p}\vert \ll \Gamma$, 
\begin{eqnarray}
&&
\partial_{t}\rho_{31} = -\left(\frac{\Gamma}{2}+i\Delta_{p}\right)\rho_{31}+\frac{i}{2}\Omega_{c}\rho_{21}+\frac{i}{2}\Omega_{p}\, ,\nonumber \\
&&
\partial_{t}\rho_{21} = i\left(\Delta_{c}-\Delta_{p}+i\gamma\right)\rho_{21}+\frac{i}{2}\Omega^{*}_{c}\rho_{31}\, ,\nonumber\\
&&
\frac{1}{c}\partial_{t}\Omega_{p}+\partial_{z}\Omega_{p}=i\eta \rho_{31}\, .\label{eq1}
\end{eqnarray}
Here, the decoherence rate $\gamma$ between the two ground states and the control (probe) laser detuning $\Delta_{c}$($\Delta_{p}$) are assumed to be negligible on the discussed short time scale $\ll\tau$. The density matrix elements $\rho_{ij}=A^{*}_{i}A_{j}$ correspond to the state vector $|\psi\rangle= A_{1}|1\rangle+A_{2}|2\rangle+A_{3}|3\rangle$. Furthermore,  $\eta$ is defined as $\eta=\frac{\Gamma\xi}{2L}$, where $\Gamma$ is the spontaneous decay rate of the excited state $|3\rangle$, $\xi$ the optical depth \cite{Lin2009, Liao2012b,Kong2014} and $L$ the target thickness. Further notations are $c$ the speed of light and $\Omega_{c(p)}$  the control (probe) Rabi frequency proportional to the corresponding laser electric field \cite{Scully2006}.

The following approximate solutions can be used to describe the dynamics with a constant $\Omega_{c}$, initial conditions $\rho_{31}(0,z)=\rho_{21}(0,z)=\Omega_{p}(0,z)=0$ and the boundary condition $\Omega_{p}(t,0)=\Omega_{p0}\delta(t-t_{0})$ such that the probe bandwidth is larger than $\Gamma$ which is in turn larger than $\Omega_{c}$ (see Supplementary Material \cite{supplmat} for derivations),
\begin{eqnarray}
\rho_{31}(T,z) &\approx&  i\frac{\Omega_{p0}}{8} J_{0}(\sqrt{\eta z T})e^{-\frac{\Gamma}{4}T}\cos\left(\frac{\Omega_{c}}{2}T\right)\, ,\label{eq4}\\
\rho_{21}(T,z) &\approx& -\frac{\Omega_{p0}}{8} J_{0}(\sqrt{\eta z T})e^{-\frac{\Gamma}{4}T}\sin\left(\frac{\Omega_{c}}{2}T\right)\, ,\label{eq5}
\end{eqnarray}
\begin{equation}
\frac{\Omega_{p}(T,z)}{\Omega_{p0}} \approx
\delta(T)
-\frac{1}{4}\sqrt{\frac{\eta z}{T}}J_{1}(\sqrt{\eta z T})e^{-\frac{\Gamma}{4}T}\cos\left(\frac{\Omega_{c}}{2}T\right)\, ,\label{eq6}
\end{equation}
where $T=t-t_{0}$ and $J_{0(1)}$ denotes the zeroth (first) order Bessel function of the first kind. 
The underlying physics for the Bessel function behavior is the dispersion of a broadband incident probe field \cite{Crisp1970,buerck1999}. Moreover, the trigonometric function dynamics results from the interference of light emission from two ATS modes \cite{Liao2012b,Autler-Townes} as long as the two ATS absorption peaks are covered by the Fourier transform of the incident broadband probe pulse, see Fig.~\ref{fig1}(c). The oscillating behavior of the coherences and of the probe field in Eqs.~(\ref{eq4}-\ref{eq6}) corresponds to a stimulated two-photon Raman process which coherently occurs on a much faster time scale than that of its spontaneous counterpart.
For a time-dependent and real $\Omega_c(t)$, Eqs.~(\ref{eq4}-\ref{eq6}) can be generalized \cite{supplmat} by introducing  $\sin\left(\frac{\Omega_c}{2}T\right)\rightarrow\sin\left[\frac{1}{2}\int_0^t\Omega_c(t')dt'\right]$ and $\cos\left(\frac{\Omega_c}{2}T\right)\rightarrow\cos\left[\frac{1}{2}\int_0^t\Omega_c(t')dt'\right]$. 
The coherent emission can be suppressed by switching off the control field when $|\rho_{21}|$ is maximal, i.e., the optical coherence completely translates into atomic coherence.
\begin{figure}[t]
\vspace{-0.4cm}
  \includegraphics[width=0.48\textwidth]{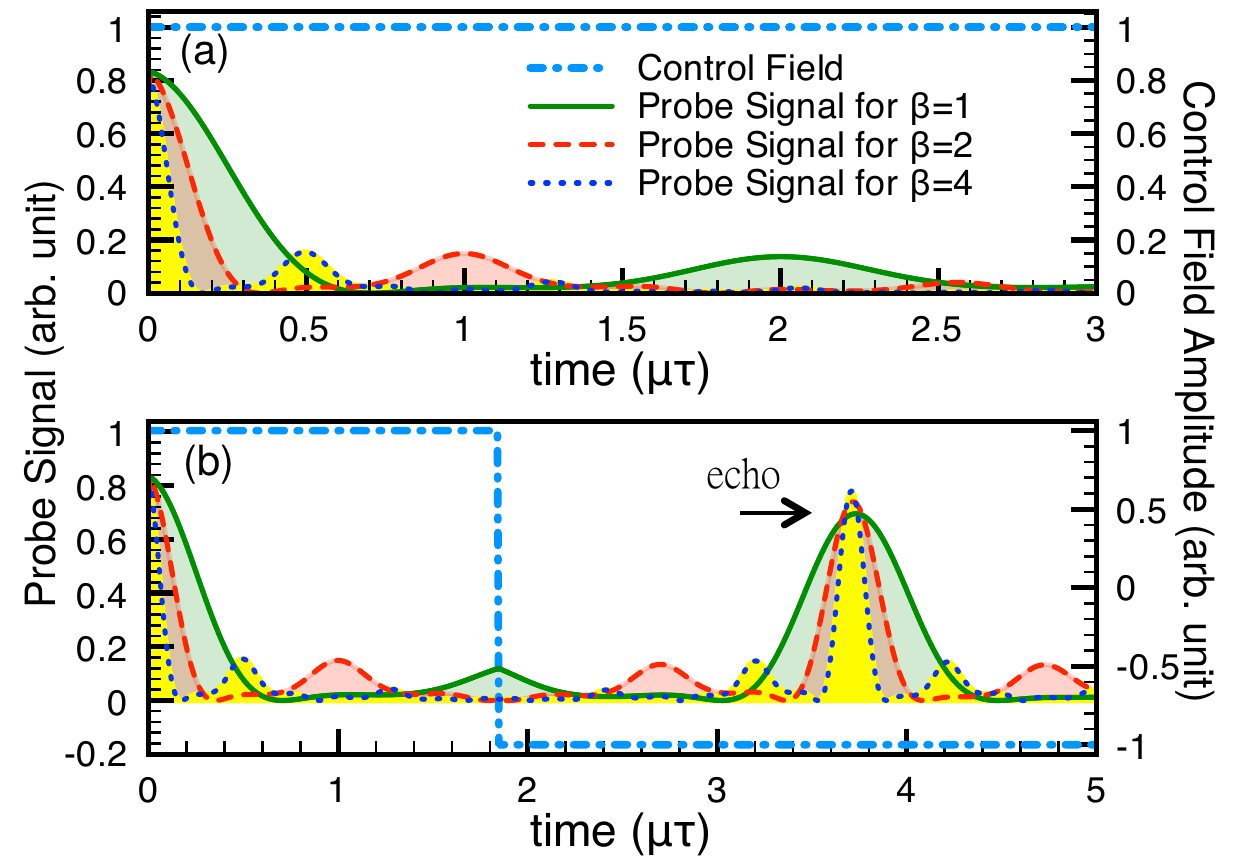}
  \caption{\label{fig2}
(Color online). (a) Probe spectra considering Gaussian control fields with different parameters $\beta$ (see text). Green solid, red dashed and blue dotted lines illustrate the scattered probe signals for $\beta=1$, 2 and 4, respectively. (b) Photon echo effect. The echoes of the scattered probe fields at around $3.6\mu\tau$ are induced by adding a $\pi$ phase shift to the control field at around $1.8\mu\tau$. All time spectra are presented in units of the lifetime $\tau$ of the excited state $|3\rangle$, e.g., $\mu\tau=10^{-6}\tau$.
  }
\end{figure}

We now turn to the all-electromagnetic photon echo effect which can be achieved with the setup illustrated in Fig.~\ref{fig1}(b). As control field we use now a Gaussian beam with position dependent Rabi frequency $\Omega_{c}(z)$ as displayed by the blue curve on the side of the green cube in Fig.~\ref{fig1}(b). Inspection of Eqs.~(\ref{eq4})-(\ref{eq6}) leads to a number of important qualitative remarks (see Supplementary Material \cite{supplmat} for derivations using a space- and time-dependent $\Omega(t,z)$):
(1) as demonstrated in Fig.~\ref{fig1}(b), a $z$-dependent control field  makes the quantum coherence evolve with an inhomogeneous rate, e.g., $\sin\left(\frac{\Omega_{c}}{2}T\right)$ becomes $\sin\left[\frac{\Omega_{c}(z)}{2}T\right]$ due to the position-dependent ATS.
(2) the projection of the spatially non-uniform ATS splitting in Fig.~\ref{fig1}(c) demonstrates that the effective linewidth of the control field modified medium is tunable by changing the profile of $\Omega(z)$. Also, an analysis of the first order scattering shows that the transmitted probe signal $|\Omega_{p}(L,T)|^{2}$ is proportional to $|\int_{0}^{L}\cos\left[\frac{\Omega_{c}(z)}{2}T\right]dz|^2$ whose duration can therefore be controlled by changing $\Omega_{c}(z)$. 
(3) a time reversal dynamics can be induced by applying a phase shift of $\pi$ to the control field, i.e., $\Omega_{c}(z)\rightarrow -\Omega_{c}(z)$,  $\sin\left[\frac{\Omega_{c}(z)}{2}T\right]\rightarrow \sin\left[-\frac{\Omega_{c}(z)}{2}T\right]$. This renders gradient echo memory (GEM) \cite{Hetet2008b,Hosseini2009,Sparkes2010,Sparkes2012,Sparkes2013} or
controlled reversible inhomogeneous broadening (CRIB) \cite{Kraus2006,Hetet2008,Tittel2010} for storing broadband pulses possible with our scheme by just changing the phase of the narrowband control field. 

To demonstrate the influence of a Gaussian beam  on the propagation of an ultra-short probe pulse, we present our Mathematica  numerical solutions (for further numerical methods details see the Supplementary Material \cite{supplmat}) of Eq.~(\ref{eq1}) in Fig.~\ref{fig2}. 
Under the assumption that the probe  spot size is much smaller than that of the control field, we can write $\Omega_{c}(z)=10^{7}\beta\Gamma/\sqrt{1+(\frac{z-L}{0.2 L})^{2}}$ with variable factor $\beta$, i.e., a $\Omega_{c}(z)$ with a Rayleigh length of $0.2 L$ are used.
We consider a medium optical thickness of  $\xi=10^{6}$ and $\Omega_{p}(T,0)=\Omega_{p0}\exp[-(\frac{T}{\kappa})^{2}]$  with $\kappa=5\times 10^{-9}\tau$ such that the broadband requirements  $\kappa^{-1}\gg\Omega_{c}\gg\Gamma$ are fulfilled.
By increasing $\beta$ from 1 to 4, the transmitted probe signals are in turn significantly compressed, as showed in Fig.~\ref{fig2}(a). This signal compression confirms our  remark (2) above, i.e., the pulse duration of the registered probe signal is determined by the maximum of $\vert\Omega_c(z)\vert$.
Furthermore, by flipping the control field at $1.8\mu\tau$, the time-inverse spectra for the probe field arise, i.e., a photon echo is generated. The echo effect can be easily understood by inspecting the spin wave picture in Fig.~\ref{fig1}(b). The green arrows depict $\rho_{21}$ at each position. By reversing the control field, $\rho_{21}$  evolves backwards and eventually becomes parallel to the initial value. Subsequently, the parallel spins  produce the echo signal that is observed at around $3.6\mu\tau$ in Fig.~\ref{fig2}(b). 

\begin{figure}[t]
\vspace{-0.4cm}
  \includegraphics[width=0.48\textwidth]{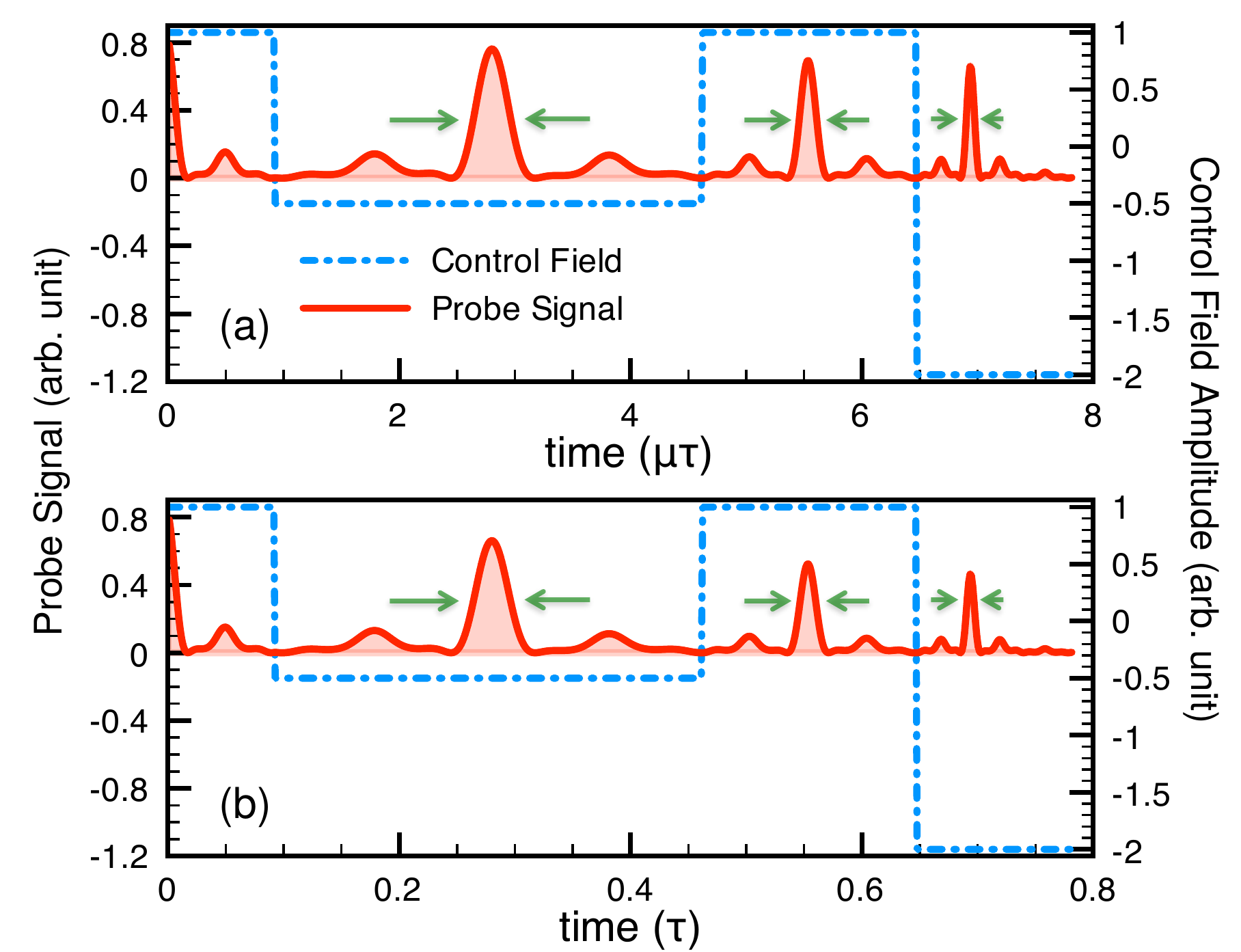}
  \caption{\label{fig3} (Color online). Modulation of probe echo signal on the time scale of (a) $\mu\tau$ and (b) $0.1 \tau$. Blue dashed-dotted lines depict control  and red solid lines  probe fields, respectively. The duration of the probe echo signals is proportional to $\vert\Omega_{c}\vert^{-1}$. Further parameters are (a)  $\xi=10^{6}$ and $\beta=4$ and (b)  $\xi=10$ and $\beta=4\times 10^{-5}$, where the values of $\beta$ correspond to control field unity on the axis.
}
\end{figure}

In what follows, we design a switching  sequence  to demonstrate that
a non-uniform $\Omega_{c}$ can control the echo signal duration for different $\beta$ values.  
In Fig.~\ref{fig3}(a), a broadband probe pulse enters the medium in the presence of $\Omega_{c}(z)$ with $\beta=4$, which causes a fast decay signal $\vert\Omega_{p}(t,L)\vert^{2}$ with a half duration of $0.05 \mu\tau$. 
Subsequently, the control field is switched to $\beta=-1$  at around $1\mu\tau$ causing an echo signal with $0.4\mu\tau$ duration to appear at around $2.8\mu\tau$. Further switches of the control field to $\beta=4$ at $4.5\mu\tau$ and  $\beta=-8$ at $6.5\mu\tau$ make the echo signal duration become $0.1\mu\tau$ and $0.05\mu\tau$, respectively. The focal $\Omega_c=\frac{\vec{P}\cdot\vec{E}_c}{\hbar}=b\Gamma$ (with $b$ constant), i.e., the control field intensity and not its bandwidth,  determines the bandwidth of the emitted echo signal. For a typical atomic dipole moment $\vert\vec{P}\vert=10^{-29}$ Cm, we estimate that a focal control field of intensity 
$c\epsilon_0\vert\vec{E}_c\vert^2=10^{-17} (b/\tau)^2$ Ws$^2$/cm$^2$ can produce an echo duration of approx.~$\tau/b$. A constraint for the echo compression arises considering the required control field modulation characterized by the rise-fall time  $t_\mathrm{rf}$. Under the condition of $\Gamma <  \Omega_c\leq	\kappa^{-1}$,  the  compressed echo duration  induced via  scaling by a factor $b^2$ in the control intensity over the time $t_\mathrm{rf}$  needs to be longer than the latter but still small compared to the (equally compressed) echo formation delay time $t_d$ such that $t_\mathrm{rf}
\leq\tau/b\leq t_d/b$.

Eqs.~(\ref{eq4})-(\ref{eq6}) further exhibit an important time scaling property when describing dynamics on different time scales.
The photon echo appears on the time scale $T\rightarrow T/s$ when $\exp[-\Gamma T/(4 s)]\approx 1$ if both the control field and medium optical thickness are switched to $\Omega_{c}\rightarrow s\Omega_{c}$ and $\xi\rightarrow s\xi$ for a fixed $L$. This provides the freedom to choose the time scale on which the echo occurs, and eventually to manipulate ultra-short laser pulses.  We demonstrate this time scaling symmetry in Fig.~\ref{fig3}(b) by using the scaled sequence of $\Omega_{c}(z)$ switches on a much longer time scale for $s=10^{-5}$, i.e., $0.1 \tau$, obtained  by choosing $\Omega_{c}$ and $\xi$ accordingly. The only qualitative difference we observe is the lower intensity of the probe after $0.4 \tau$ caused by spontaneous decay.

\begin{figure}[t]
\vspace{-0.4cm}
  \includegraphics[width=0.48\textwidth]{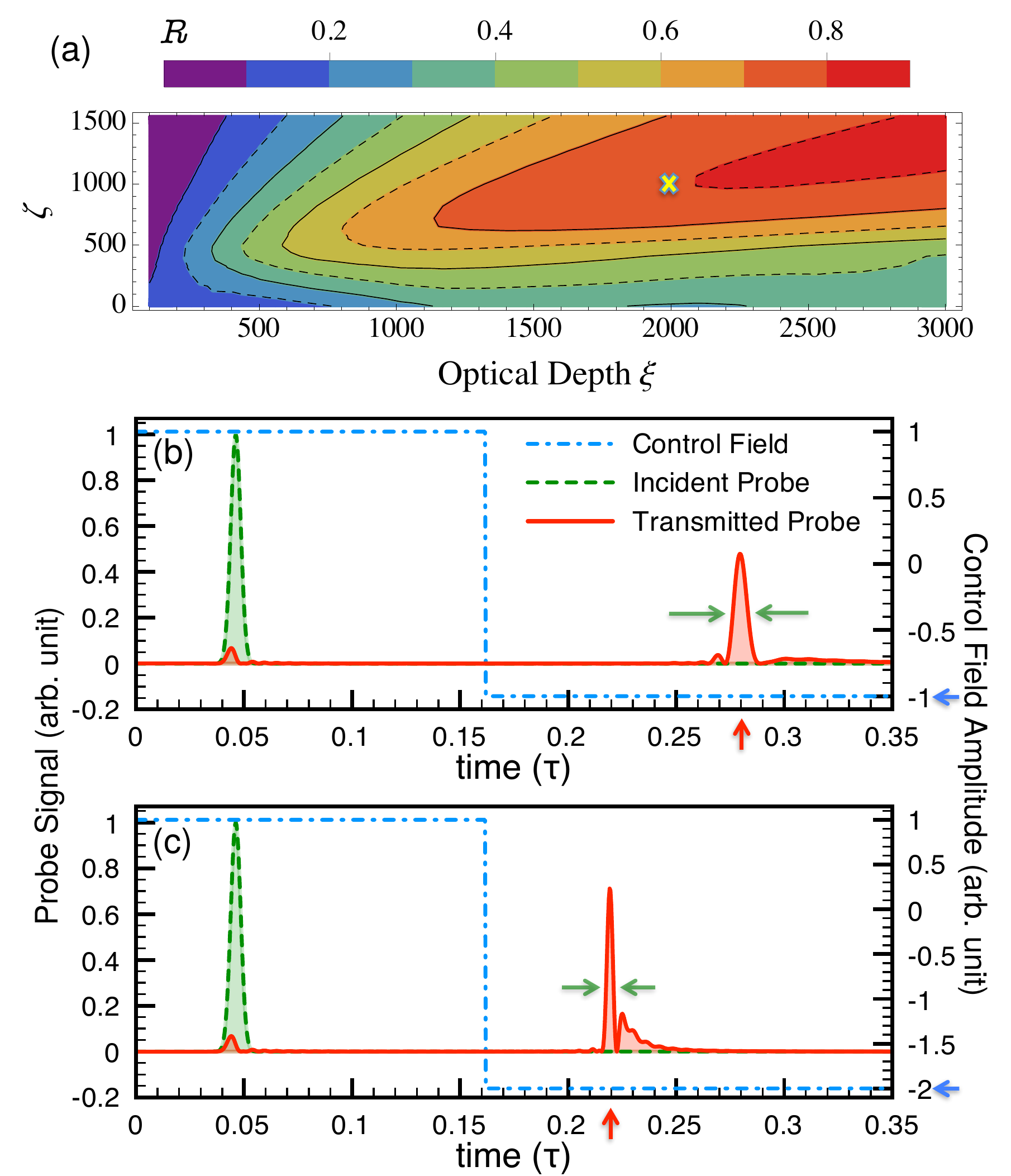}
  \caption{\label{fig4} (Color online). (a) Contour plot of the storage efficiency $R(\xi,\zeta)$ for a linear control field intensity $\Omega_{c}(z)=\zeta\Gamma z/L$. 
  Yellow cross indicates the parameter set $(\xi,\zeta)=(2000,1000)$ used in (b,c) to demonstrate  storage, retrieval and generation of a broadband pulse via photon echo. 
  (b) around 80\% of the incident  probe (green dashed line) with a duration of $\kappa=5\times 10^{-3}\tau$ are stored as target quantum coherence. By switching $\Omega_{c}\rightarrow -\Omega_{c}$ at $0.16 \tau$, an echo signal (red solid line) with a duration of $5\times 10^{-3}\tau$ is emitted. 
  (c) switching $\Omega_{c}\rightarrow-2\Omega_{c}$ (blue dashed-dotted line)  releases a shorter echo signal with a duration of  $2.5\times 10^{-3}\tau$. Red arrows on the abscissa and blue arrows on the ordinate indicate  the echo peak times and the control field intensity,  respectively.
}
\end{figure}

Finally, we focus on the question whether a broadband light pulse, i.e., $\kappa^{-1}\gg\Gamma$, can be stored using our scheme. 
For a quantum memory device we envisage  a linear $\Omega_{c}(z)=\zeta\Gamma z/L$ to equally distribute each frequency along the target \cite{Sparkes2010}. A probe pulse with $\kappa=0.005\tau$ (bandwidth of $200\Gamma$) impinges on a target at $t=0.048\tau$. Subsequently, the stored pulse is retrieved as an echo signal by applying a $\pi$ phase shift to $\Omega_{c}$ at $t=0.065\tau$, i.e., right after probe field's complete entrance into the target.
The calculated total storage efficiency $R(\xi,\zeta)=\int_{0.065\tau}^{\infty}\vert\Omega_{p}(t,L)\vert^{2}dt/\int_{0}^{\infty}\vert\Omega_{p}(t,0)\vert^{2}dt$ is illustrated in Fig.~\ref{fig4}(a). High storage efficiency requires an optically thick medium. To obtain a storage efficiency higher than 80\%, a target with $\xi\geq 2000$ is required. The dispersion of the echo signal is negligible in the domain $\zeta >\xi$, becoming however visible for $\zeta\leq\xi$. 
Figure~\ref{fig4}(b) shows a case of moderate echo distortion using the parameters indicated by the yellow cross in Fig.~\ref{fig4}(a). Flipping the control field at $t=0.16\tau$ leads to the generation of an echo at $t=0.28\tau$ with the same pulse duration and with a classical fidelity \cite{Chen2013} of 75\%, suggesting the possibility of storing a broadband light pulse with $\kappa^{-1}\gg\Gamma$ via our scheme. Furthermore, generating an echo signal with a broader bandwidth than that of the incident pulse is showed in Fig.~\ref{fig4}(c) by changing the control field  $\Omega_c \rightarrow -2\Omega_{c}$. An echo signal with a shorter pulse duration of $0.0025\tau$ is emitted at an earlier time $t=0.22\tau$.

A comparison with EIT-based methods \cite{Chen2013} is not straight-forward  since the two schemes typically address different parameter regimes.
With the stored pulse bandwidth restricted by the maximum $\vert\Omega_c(z)\vert$ for both cases, the control laser power consumption is expected to be smaller in our scheme since the maximum $\vert\Omega_c(z)\vert$ occurs only at the focus. 
Our scheme focuses on the broadband excitation regime where EIT-based methods do not reach the optimal retrieval efficiency of  $R_r= 1-2.9/\xi$  \cite{Gorshkov2007b}.
For fixed storage bandwidth, i.e., in our case the maximum $\Omega_c$, and fixed values of $R$, one could compare the required $\xi$ for each scheme. 
Considering  $R=75$\%, a spatially uniform $\Omega_c=1000\Gamma$ for EIT and the parameters used in Fig.~\ref{fig4}(b), EIT requires an optical thickness of  $\xi=10000$, while the corresponding value for our scheme is  $\xi=2000$. In turn, for fixed $\Omega_c$ and $\xi$ values, the fidelity (around 61\%) and the delay-bandwidth product \cite{Jonas2007,Baba2008} reached in an EIT scheme are smaller than those of our setup. Since in addition the echo 
delay time can be freely chosen,  our storage scheme for controlling
broadband excitations could become competitive and even present advantages such as more  flexible buffering capacity compared to EIT-based slow light setups.

Fast echo control
requires an optically thick medium together with  focusing or a beam shaper for the control field. A very large optical thickness can be achieved in nuclear systems, where a concentration of $10^{18}/{\rm cm}^3$ of doped $^{229}\mathrm{Th}$ nuclei in a vacuum ultraviolet-transparent crystal \cite{kazakov2012} leads to an optical thickness of up to $\xi=10^{6}$ \cite{Liao2012b}. For more typical systems in the optical regime, an optical depth $\xi$ over 1000 has already been experimentally achieved in cold atom systems \cite{Blatt2014}, e.g. cold $^{87}$Rb gas in a two-dimensional magneto-optical trap \cite{Sparkes2013}. The typical medium length $L$ of 1--5 cm \cite{Sparkes2013,Chen2013,Shiau2011,kazakov2012} restricts the spatial profile of the control laser. We estimate the required Rayleigh length $r$ and the power $K$ of the Gaussian beam via the expressions $b\Gamma/\sqrt{1+(L/r)^2}=\Gamma$ and $K=\frac{1}{2}c\epsilon_0 \vert\vec{E}_c\vert^2 \lambda r$, where $\lambda$ is the laser wavelength, and $b\Gamma$ is the maximum Rabi 
frequency of the control field, i.e.,  the maximum bandwidth of the photon emission. Considering the storage results in Fig.~\ref{fig4}(b) with a cold $^{87}$Rb atomic gas, $L=5$ cm, $\lambda$ of 780 nm or 795 nm and $b=1000$, our scheme  requires $r=50$ $\mu$m with a laser focusing of a 0.08 W cw laser on a spot size of 40 $\mu$m$^2$. Alternatively, for $b<100$, one could use a perpendicular setup \cite{Hau1999} that rotates the control laser with 90 degrees such that the laser gradient along the medium can be adjusted by changing the transverse laser profile. The required focus spot size $\pi w^2$ can be estimated from $b\Gamma\exp[-(L/w)^2]=\Gamma$, where $w$ is the Gaussian beam waist size,  resulting in a focusing of a few-kW laser on a spot size of $\pi L^2/\ln(b)=10^9 \mu$m$^2$ for the optical parameters above.
These are standard parameters in laser experiments at present being far away from the optical diffraction limit.  Moreover, a high efficient optical beam shaper composed of, e.g., liquid-crystal spatial light modulators, phase plates and deformable mirrors has been recently put forward \cite{Sparkes2010}. Once experimentally realized, our scheme  may not only ease the need of a broadband read-write field for light storage at large bandwidths \cite{deRiedmatten2010} but also allow for  flexible manipulation of broadband excitations and light pulses on different time scales.

The authors would like to thank T. Pfeifer, Y.-H. Chen and W. Hung for helpful discussions about the experimental feasibility.
 
\bibliographystyle{apsrev}
\bibliography{echo}
\end{document}